\documentclass[preprint,showpacs,preprintnumbers,amsmath,amssymb,prb]{revtex4}

\usepackage{graphicx}
\usepackage{dcolumn}
\usepackage{bm}

\begin{document}


\title{Higher order and infinite Trotter--number extrapolations in 
path integral Monte Carlo}

\author{L. Brualla}
\affiliation{%
International School for Advanced Studies, SISSA, and\\
INFM DEMOCRITOS National Simulation Center\\
Via Beirut, 2--4, I--34014 Trieste, Italy
}%
\author{K. Sakkos}
\author{J. Boronat}%
\author{J. Casulleras}
\affiliation{
Departament de F\'{\i}sica i Enginyeria Nuclear,\\
Universitat Polit{\`e}cnica de Catalunya,\\
Campus Nord B4--B5, E--08034 Barcelona, Spain
}%

\date{\today}

\begin{abstract}
Improvements beyond the primitive approximation in the path integral Monte
Carlo method are explored both in a model problem and in real systems. Two
different strategies are studied: the Richardson extrapolation on top of the path
integral Monte Carlo data and the Takahashi-Imada action. The Richardson
extrapolation, mainly combined with the primitive action, always reduces the
number-of-beads dependence, helps in determining the approach to the
dominant power law behavior, and all without additional computational cost. The
Takahashi-Imada action has been tested in two hard-core interacting quantum
liquids at low temperature. The results obtained show that the fourth-order
behavior near the asymptote is conserved, and that the use of this improved
action reduces the computing time with respect to the primitive
approximation.
\end{abstract}

\maketitle

\section{\label{sec:introduction}Introduction}
Path integral Monte Carlo (PIMC) has become in the last decade a standard
tool for studying quantum liquids and solids at finite
temperature.\cite{gillan,ceperley,chakra} The
PIMC method allows for the calculation of the quantum-statistical partition
function, and from it, thermodynamic functions like the internal energy or
the specific heat. The basis on which PIMC rests is the analytical continuation 
to imaginary time of Feynman's path-integral formalism of Quantum
Mechanics.\cite{fey1,fey2,kleinert} The resulting convolutive
property for the thermal density matrix makes accessible the quantum physics
of low temperatures starting from density matrices at higher temperatures.        
At sufficiently high temperature, commutator terms between the kinetic and
potential operators defining the Hamiltonian of the system can be
disregarded, and then, the density matrix factorizes as a product of the
kinetic and potential parts. This is known as the primitive approximation
(PA) and the unbiased convergence to a desired lower temperature by
increasing the number $M$ of convolutive terms is guaranteed by the Trotter
formula.\cite{trotter} 

The expectation value of an operator through the thermal density matrix is
written in this language as a multidimensional integral appropriate for a
Monte Carlo interpretation. The quantum problem can be mapped to a
classical one of interacting polymers, each one representing an
atom.\cite{barker,chandler} In
the action, beads of different atoms but with the same index interact 
through the potential, and
all the beads of the same atom form a closed chain of springs. At the
level of the PA and avoiding the specific quantum statistics of the atoms
or molecules, relevant at temperatures near the absolute zero, the
PIMC method looks rather simple. However, in the implementation of the
algorithm there are technical aspects which require of some additional
effort. The main point is that, as the number of beads of the polymeric chains 
increases
rapidly with decreasing temperature, the sampling of the probability
distribution function becomes almost impossible by means of individual
movements.
Several mechanisms for performing smart collective movements
of the beads have been proposed and proved to be able of eliminating the
slowing down present in the single-bead schemes. Worth mentioning are the
staging\cite{pollock,sprik,tucker} and bisection\cite{ceperley,gordillo}
 which exploit the fact that the free action (kinetic
part in an interacting system) can be sampled by non-rejection methods, and
therefore, only the potential part requires a Metropolis step. In a large
number of PIMC simulations, these collective sampling schemes are enough
to reach a reasonable degree of efficiency. Nevertheless, calculations of
quantum fluids at extremely low temperature require chains composed by
several thousands of beads, feature which makes the problem more involved.      

The number of beads required to achieve the asymptotic regime, at a given
temperature, depends on the approximation taken for the action. 
It has been proved that PA is accurate to order $(\beta/M)^2$, with $\beta$ 
the inverse temperature and $M$ the number of beads.\cite{raedt,fye} 
Focusing the analysis
on PIMC, and thus not considering the specific developments for the Fourier
PIMC method,\cite{doll} the improvement of the action can be pursued following at
least two different alternatives. In the first one, the PA is substituted by
the pair-product action in which the basic step on the chain is constructed as the
exact action for two isolated atoms.\cite{pollock} This method is more
efficient than the PA, mainly when dealing with hard-sphere-like systems,
and it is the only method up to date which has proved to be accurate enough 
to reach the superfluid regime of liquid $^4$He.\cite{ceperley} The second possibility
relies on the inclusion in the action of the lowest-order corrections to
the primitive approximation for the exponential of the Hamiltonian $e^{-
\beta \widehat{H}}$. To this end, Takahashi and Imada\cite{taka} and later on, and
independently, Li and Broughton\cite{li} found a manageable expression for the
trace accurate to order $(\beta/M)^4$. In this approximation,
hereafter referred as TIA, the beads of different atoms interact through an
effective potential which is the sum of the interatomic potential and a
$\beta$-dependent term containing the double commutator $[[V,K],V]$, with
$V$ and $K$ the potential and kinetic operators, respectively.  Recently,
Jang \textit{et al.}\cite{jang} have used in PIMC a genuine fourth-order factorization
according to the proposal of Suzuki\cite{suzuki} and the posterior
developments of Chin.\cite{chin1,chin2}
However, the accuracy achieved in several test problems by Jang \textit{et
al.}\cite{jang} is comparable to the one reached using the simpler TIA.

In the present paper we analyze the $\beta/M$ dependence of the energy using
both PA and TIA for a test model and for two strongly interacting fluids at
low temperature, i.e., Ne and $^4$He. In the analysis of the PA results, we
introduce the Richardson extrapolation.\cite{recipes} The efficiency of this
numerical scheme for extrapolation is well known in fields like
integration or solution of differential equations. We show that its use in
PIMC can also be useful in two directions. First, its simple use helps
in determining when the expected law $(\beta/M)^2$ is reached; second, the
successive extrapolations to $M \rightarrow \infty$ follow a nearly
fourth-order dependence and then they always approach faster to the
asymptote than the PA. The second main objective of the work is the study
of the TIA when applied to hard-core interacting systems in the regime of
low temperatures since the applications of TIA to these situations is
rather scarce.\cite{muser,lacks,weht,singer} Previous tests of the TIA to systems not so dense and not so
cold have shown the expected fourth-order accuracy.\cite{muser} In fact, the
achievement of a fourth-order law for hard-core potentials using TIA has
been some times questioned\cite{ceperley} but never thoroughly analyzed. Our present
results show that in the regime studied the accuracy obtained is the
expected one and the global efficiency of TIA is larger than the one of the
standard PA. On the other hand, the use of
the Richardson extrapolation on top of the TIA results has proved not to be
so helpful, at least in the temperatures where the two real systems have
been studied. 


The rest of the paper is organized as follows: Section II contains the formalism of the
PA and the TIA, including the expressions for the thermodynamic and
centroid-virial estimators for the energy. The application of the
Richardson extrapolation to the PIMC calculations is also discussed. In
Sec. III, results for the three systems under study are presented with
special emphasis to the $(\beta/M)$ dependence of the energy obtained using the
two models for the action. Finally, Sec. IV comprises a brief summary and
the main conclusions of the work.

\section{\label{sec:lbtheory} Formalism}
The partition function of a system described by a Hamiltonian $\widehat{H}$ at a
temperature $T=1/\beta$ is
\begin{equation}
Z=\textrm{Tr}\  e^{-\beta \widehat{H}}    \ .
\label{partition}
\end{equation}
We consider a system composed by $N$ particles with $\widehat{H}=\widehat{K}+\widehat{V}$,
the kinetic operator being
\begin{equation}
\widehat{K}=-\frac{\hbar^2}{2m} \, \sum_{i=1}^{N} \bm{\nabla}_i^2 \ ,
\label{opkin}
\end{equation}  
and the potential one 
\begin{equation}
\widehat{V} = \sum_{i<j}^{N} V(r_{ij}) \ ,
\label{oppot}
\end{equation}
where pairwise interactions are assumed. The PA is the lowest order term in
powers of $(\beta/M)$ of the exponential of the sum of the two operators
$\widehat{K}$ and $\widehat{V}$,
\begin{equation}
\textrm{Tr}\ e^{-\beta (\widehat{K}+\widehat{V})} \simeq 
\textrm{Tr}\ \left( e^{-(\beta \widehat{K}/M)} e^{-(\beta \widehat{V}/M)}
\right)^M \ .
\label{splitpa}
\end{equation}
In fact, in the limit where $M \rightarrow \infty$ the previous expression
becomes exact, yielding the Trotter formula.
With the decomposition (\ref{splitpa}), and restricting the analysis to
distinguishable particles, the partition function becomes
\begin{equation}
Z= \int d\bm{R}_1 \ldots d\bm{R}_M \ \prod_{\alpha=1}^{M}
\rho_{\textrm{PA}}(\bm{R}_\alpha,\bm{R}_{\alpha+1}) \ ,
\label{zpa1}
\end{equation}
with $\bm{R}\equiv \{\bm{r}_1,\ldots,\bm{r}_N \}$ and
$\bm{R}_{M+1}=\bm{R}_1$. Greek and Latin indexes are used throughout the
paper to denote beads and particles, respectively. 
The primitive factorization (\ref{splitpa}),
which decouples the kinetic and potential operators, leads to the well-known
PA,
\begin{equation}
\rho_{\textrm{PA}}(\bm{R}_\alpha,\bm{R}_{\alpha+1}) = 
\left( \frac{M m}{2 \pi \beta \hbar^2} \right)^{3N/2} \exp \left\{
-\sum_{i=1}^{N} \frac{M m}{2 \beta \hbar^2}
(\bm{r}_{\alpha,i}-\bm{r}_{\alpha+1,i})^2 -\frac{\beta}{M} \sum_{i<j}^{N}
V(r_{\alpha,ij}) \right\}   \ .
\label{zpa2}
\end{equation} 

Takahashi and Imada\cite{taka} proved that it is possible to extend the
accuracy of the action up to fourth order for the trace by substituting in
Eq. (\ref{splitpa}) the operator $\widehat{V}$ by another one $\widehat{W}$
given by
\begin{equation}
\widehat{W} = \sum_{i<j}^{N} V(r_{ij}) + \frac{1}{24} \frac{\hbar^2}{m} 
\left( \frac{\beta}{M} \right)^2 \sum_{i=1}^{N} |\bm{F}_{i}|^2 \ .
\label{potti1}
\end{equation}
The second term on the r.h.s of Eq. (\ref{potti1}) corresponds to the
double commutator $[[V,K],V]$ appearing in the development of the
exponential $e^{-\beta \widehat{H}}$. The classical-like force $\bm{F}_i$
is defined as
\begin{equation}
\bm{F}_i = \sum_{j \neq i}^{N} \bm{\nabla}_i V(r_{ij}) \ .
\label{forcelb}
\end{equation}
Therefore, the partition function using the TIA is written like the PA one
(\ref{zpa1}) just substituting the interatomic potential $V(r_{\alpha,ij})$ in Eq.
(\ref{zpa2}) by $W(r_{\alpha,ij})$,
\begin{eqnarray}
\rho_{\textrm{TIA}}(\bm{R}_\alpha,\bm{R}_{\alpha+1}) & = &
\left( \frac{M m}{2 \pi \beta \hbar^2} \right)^{3N/2} \exp \left\{
-\sum_{i=1}^{N} \frac{M m}{2 \beta \hbar^2} 
(\bm{r}_{\alpha,i}-\bm{r}_{\alpha+1,i})^2   \right.  \\ \nonumber
& & \left.  -\frac{\beta}{M} \left( \sum_{i<j}^{N} V(r_{\alpha,ij}) 
+ \frac{1}{24} \frac{\hbar^2}{m} 
\left( \frac{\beta}{M} \right)^2 \sum_{i=1}^{N} |\bm{F}_{\alpha,i}|^2 \right)
\right\}  \ .
\label{rhotia}
\end{eqnarray}

First derivatives of the partition function allow for the estimation of the
total and partial energies of the $N$-body system,\cite{fey2}
\begin{eqnarray}
\frac{E}{N} & = &  - \frac{1}{N Z} \, \frac{\partial Z}{\partial \beta}  
  \label{totener} \\
\frac{K}{N} & = & \frac{m}{N \beta Z} \, \frac{\partial Z}{\partial m}
\label{kinener} \, 
\end{eqnarray}
and $V/N=E/N-K/N$. The potential energy, and any other coordinate
operators, can also be obtained through the general relation\cite{taka}
\begin{equation}
O(\bm{R}) = -\frac{1}{\beta} \frac{1}{Z(V)} \left. \frac{d Z(V+\lambda
O)}{d \lambda} \right|_{\lambda=0}   \ .
\label{nice}
\end{equation}
Equation (\ref{kinener}) leads to the thermodynamic estimators for the
kinetic energy. At the level of PA,
\begin{equation}
\frac{K_{\textrm{PA}}^{\textrm{th}}}{N} = \frac{3 M}{2  \beta} -
\sum_{\alpha=1}^{M} \sum_{i=1}^{N} \frac{M m}{2 N \beta^2 \hbar^2}
(\bm{r}_{\alpha,i} -\bm{r}_{\alpha+1,i})^2 \ .
\label{thermopa}
\end{equation} 
In the case of the TIA the effective potential $W$ depends on the mass of
the particles and the temperature. Therefore, explicit terms depending on
the interatomic potential appear in the estimation of the kinetic energy.
Explicitly,
\begin{equation}
\frac{K_{\textrm{TIA}}^{\textrm{th}}}{N} = \frac{K_{\textrm{PA}}^{\textrm{th}}}{N}
+\frac{1}{24}\, \frac{\hbar^2}{m} \, \frac{\beta^2}{ N M^3} \sum_{\alpha=1}^{M}
\sum_{i=1}^{N} |\bm{F}_{\alpha,i}|^2 \ .
\label{thermotia}
\end{equation}
The potential energy in the PA case corresponds to a direct estimation of
$V(r)$,
\begin{equation}
\frac{V_{\textrm{PA}}}{N} = \frac{1}{NM} \, \sum_{\alpha=1}^{M}
\sum_{i<j}^{N} V(r_{\alpha,ij}) \ ,
\label{potpa}
\end{equation}
whereas the application of Eq. (\ref{nice}) for the TIA produces the same
additional force term of Eq. (\ref{thermotia}), multiplied by a factor of
two, 
\begin{equation}
\frac{V_{\textrm{TIA}}}{N} = \frac{V_{\textrm{PA}}}{N} 
+\frac{1}{12}\, \frac{\hbar^2}{m} \, \frac{\beta^2}{ N M^3} \sum_{\alpha=1}^{M}
\sum_{i=1}^{N} |\bm{F}_{\alpha,i}|^2  \ .
\label{pottia}
\end{equation}

The unbiased convergence of both the PA and TIA factorizations to the exact
energy when $M$ increases is granted by the Trotter formula.\cite{trotter} However, the
thermodynamic estimator for both approaches (\ref{thermopa},\ref{thermotia})
presents the drawback of a statistical variance which increases with the
number of beads $M$.\cite{herman,janke} To overcome this problem one can derive another
estimator for the energy by exploiting the invariance of the partition
function under a rescaling of the particle positions ($\bm{r}_i \rightarrow
\lambda \bm{r}_i$). This leads to the so called virial
estimator\cite{herman} in which  the
kinetic energy is obtained from derivatives of the interatomic potential
$V(r)$. It has been proved that, contrary to the thermodynamic estimator,
 the variance of the virial estimator is roughly constant as a function of
 $M$.\cite{herman,janke} We have chosen in the present work the centroid version of the
 virial estimator. In the PA,
\begin{equation}
\frac{K_{\textrm{PA}}^{\textrm{cv}}}{N} = \frac{3}{2 \beta} + \frac{1}{2}
\, \frac{1}{N M} \sum_{\alpha=1}^{M} \sum_{i=1}^{N} (\bm{r}_{\alpha,i} -
\bm{r}_{0,i} ) \cdot \bm{F}_{\alpha,i} \ ,  
\label{virialpa}
\end{equation}   
with $\bm{r}_{0,i}=(\bm{r}_{1,i}+\ldots+\bm{r}_{\alpha,i})/M$ the centroid
position of atom $i$. By applying the same method to the TIA, two new terms
appear requiring up to the second derivative of $V(r)$,
\begin{equation}
\frac{K_{\textrm{TIA}}^{\textrm{cv}}}{N} = \frac{K_{\textrm{PA}}^{\textrm{cv}}}{N}
+\frac{1}{24}\, \frac{\hbar^2}{m} \, \frac{\beta^2}{ N M^3} \sum_{\alpha=1}^{M}
\sum_{i=1}^{N}  \left\{       |\bm{F}_{\alpha,i}|^2
+ \sum_{j \neq i}^{N} (r_{\alpha,i} - r_{0,i} )^a \, T(\alpha,i,j)^b_a \, 
(F_{\alpha,i} - F_{\alpha,j})_b \right\} \ ,
\label{virialtia} 
\end{equation}
where $\{a,b\}$ stand for the Cartesian coordinates and an implicit
summation over repeated indices has been assumed. The tensor $T$ is
explicitly written as
\begin{equation}
T(\alpha,i,j)^b_a = \left[ \frac{\delta_a^b}{r_{\alpha,ij}} -
\frac{(r_{\alpha,ij})^b (r_{\alpha,ij})_a}{ r_{\alpha,ij}^3} \right] \frac{d
V(r_{\alpha,ij})}{d r_{\alpha,ij}} + 
\frac{(r_{\alpha,ij})^b (r_{\alpha,ij})_a}{ r_{\alpha,ij}^2} \, \frac{d^2
V(r_{\alpha,ij})}{d r_{\alpha,ij}^2} \ .
\label{tensortia}
\end{equation}

The PIMC method is the most efficient theoretical tool to deal with a
microscopic description of systems at low temperatures where quantum
effects are unavoidable. With the only external assumption of the
interatomic potentials, PIMC provides exact results at a given temperature
with a self-contained adjustment of the number of terms (beads $M$) in the
action to the \textit{quanticity} of the system: from $M=1$, which
corresponds to classical Monte Carlo (each particle is a point in the
configuration space), to values of $M$ (each particle is represented
by a closed chain of beads) larger enough to fulfill the Trotter
formula.\cite{trotter}
It is hence essential to remove from the calculation the bias coming from
the use of a discrete value for $M$, or at least to reduce it to the level
of the typical statistical error. In order to help in this analysis we have
used the well-known Richardson extrapolation;\cite{recipes} its use in numerical
integration, derivation or differential equations permits to achieve high
accuracy by using low-order formulas.    

The $M$-dependence of integrated quantities like the energy for the PA and
TIA, when $M \rightarrow \infty$, is of the form
\begin{equation}
E = E_0 + A_{\delta} \left( \frac{1}{M} \right)^{\delta} + 
A_{\delta+2} \left( \frac{1}{M} \right)^{\delta+2} + \ldots    \ ,
\label{asympt}
\end{equation} 
with $\delta=2,4$ for PA and TIA, respectively. The value $1/M$ in PIMC
plays the same role than the step size $h$ in a numerical integration and in
both cases one is interested in the extrapolation to the \textit{ideal} case
$h=0$ ($1/M=0$). Richardson extrapolation is a simple and clever way of performing
that extrapolation improving the order of the approach. In the present
case, the extrapolation to $M=\infty$ is given by
\begin{equation}
  E_{\infty}^{(\delta)} = E_2 + \left\{ \frac{(M_1/M_2)^\delta}{1-(M_1/M_2)^\delta} \right\}
               (E_2 - E_1) \ ,
\label{richardsoninf}
\end{equation} 
$E_1$  and $E_2$ being the energies estimated using $M_1$ and $M_2$ beads,
respectively ($M_2>M_1$). When the calculation proceeds, and  $M$ is
progressively increased it is also useful to know how far one is from the
asymptotic law (\ref{asympt}). In this respect, the Richardson
extrapolation can also be used to predict the energy for a number of beads 
$M>M_2>M_1$ through
\begin{equation}
      E_M^{(\delta)} = E_{1} + 
      \left\{ \frac{1 - (M_1/M)^\delta}{1 - (M_1/M_2)^\delta} \right\} 
      \left( E_{2} - E_{1} \right)
      \  .
\label{richardtwopoint}
\end{equation}

\section{Results}
The usefulness of the Richardson extrapolation on top of the PIMC
calculations and the accuracy of the TIA with respect to the PA have been
studied in three different systems. The first one corresponds to the test
problem of a particle in a one-dimensional harmonic potential in which the
exact solution is known. In the second one, we study liquid Ne at 25.8 K, 
a real system at relatively high density and where the interatomic
interaction presents a hard core at short distances. Finally, the most
exigent test in the present analysis corresponds to the calculation of the
energy of liquid $^4$He at 5.1 K in which the low temperature, the small
mass, and the hard core of the interactions make the quantum effects much
larger. 

The sampling of the three systems studied has been carried out by combining collective
movements of some beads of a given chain and movements of the center of
mass of each one of the chains representing the atoms. In both movements
the size of the proposed movements is fixed to keep a Metropolis acceptance 
of 30-50\%. As it has been proved, the introduction of multibead movements
is unavoidable when the number of beads increases since they
eliminate the slowing down observed in a bead by bead sampling. To this end,
we have used the bisection\cite{ceperley,gordillo} and
staging\cite{pollock,sprik,tucker} methods: the bisection up to level
three for short chains and staging for longer ones. Both methods correspond
to an exact sampling of the free (kinetic) part of the action and
therefore,
in the Metropolis step, only the potential part of the action is sampled.

\subsection{\label{sec:harmonic}Harmonic oscillator}

We consider a particle in a one-dimensional harmonic well with Hamiltonian
\begin{equation}
       \widehat{H}=-\frac{\hbar ^2}{2m}
       \frac{\partial ^2}{\partial x^2}+ \frac{1}{2}\, m \omega ^2 x^2 \ .
\label{hamiltonianharmonicoscillator}
\end{equation}
In this problem, the partition function and the energy are exactly
known,\cite{kleinert}
\begin{eqnarray}
Z & = &  \left[ 2 \sinh(\beta \hbar \omega / 2 ) \right]^{-1} 
\label{zharmo} \\
       E & = & 
       \frac{1}{2} \hbar \omega \coth 
       \left( \beta \hbar \omega / 2 \right) \ .
\label{TheoreticalHarmonicEn}       
\end{eqnarray}
In the  PIMC calculation we have assumed $\omega = \hbar = m =1$, and $T=0.2$.  
The exact energy from Eq. (\ref{TheoreticalHarmonicEn}) at this temperature 
and using these reduced units is 0.50678.

\begin{table}[b]
\centering
\begin{ruledtabular}
\begin{tabular}{ccccccc}
 $M$  & $E_{\text{PA}}$ & $E_{M}^{(2)}$ & $E_{\infty}^{(2)}$ &
 $E_{\text{TIA}}$ &  $E_{M}^{(4)}$ & $E_{\infty}^{(4)}$ \\   
\hline 
2   &  0.30755   &         &         & 0.44702 &          &       \\
4   &  0.43162   &         &         & 0.50053 &	  &       \\
8   &  0.48424   & 0.46264 & 0.47298 & 0.50630 &  0.50387 & 0.50410    \\
16  &  0.50085   & 0.49740 & 0.50178 & 0.50675 &  0.50666 & 0.50668  \\
32  &  0.50528   & 0.50500 & 0.50639 & 0.50678 &  0.50678 & 0.50678  \\
64  &  0.50641   & 0.50639 & 0.50676 & 0.50678 &  0.50678 & 0.50678	  \\
128 &  0.50669   & 0.50669 & 0.50679 &         &  0.50678 & 0.50678            \\
256 &  0.50676   & 0.50676 & 0.50678 &         &	  &	    \\
512 &  0.50678   & 0.50678 & 0.50678 &         &	  &	     \\ 	   
1024 &           & 0.50678 & 0.50679 &	       & 	 &	   \\
\end{tabular}
\end{ruledtabular}
\caption{PA ($E_{\text{PA}}$) and TIA ($E_{\text{TIA}}$) results for the 
one-dimensional harmonic oscillator at
$T=0.2$. Richardson extrapolations to $M$ ($E_{M}^{(\delta)}$) and to
$M\rightarrow \infty$ ($E_{\infty}^{(\delta)}$), using the $M/4$ and
$M/2$ energies, are also reported
($\delta=2,4$ stands for PA and TIA, respectively).}
\label{Richardsonextrapolationho}
\end{table}

In Table \ref{Richardsonextrapolationho}, PIMC results obtained using the
PA and TIA approximations are reported as a function of the number of beads
$M$. As already known for this model problem,\cite{taka,li} the accuracy of
$E_{\text{TIA}}$ is manifestly superior to the PA energies $E_{\text{PA}}$.
Within a statistical fluctuation of $10^{-5}$, PA
arrives to the exact energy when $M=512$ whereas TIA requires only $M=32$,
a figure sixteen times smaller. In the same Table, we report Richardson
extrapolations to a given $M$ ($E_{M}^{(\delta)}$, Eq. \ref{richardtwopoint}) 
and to 
$M\rightarrow \infty$  ($E_{\infty}^{(\delta)}$, Eq. \ref{richardsoninf}), 
derived from calculated results ($E_{\text{PA}}$, $E_{\text{TIA}}$)
using $M/4$ and $M/2$ beads. 
When $\delta=2$, i.e., in the PA, the extrapolation to $M \rightarrow
\infty$ achieves the asymptote with only $M=64$, a factor of eight smaller
than using only the direct output  $E_{\text{PA}}$. The extrapolation to
a next $M$ is also helpful to know when the expected quadratic law is
attained since there that extrapolation coincides with the direct PA
calculation. Finally, in  Table I, the Richardson
extrapolations from  TIA results are shown. In this case, the gain achieved by the
extrapolation is significantly smaller since only a factor of two seems
reachable. 

\begin{figure}
   \centerline{\includegraphics[width=0.7\linewidth]{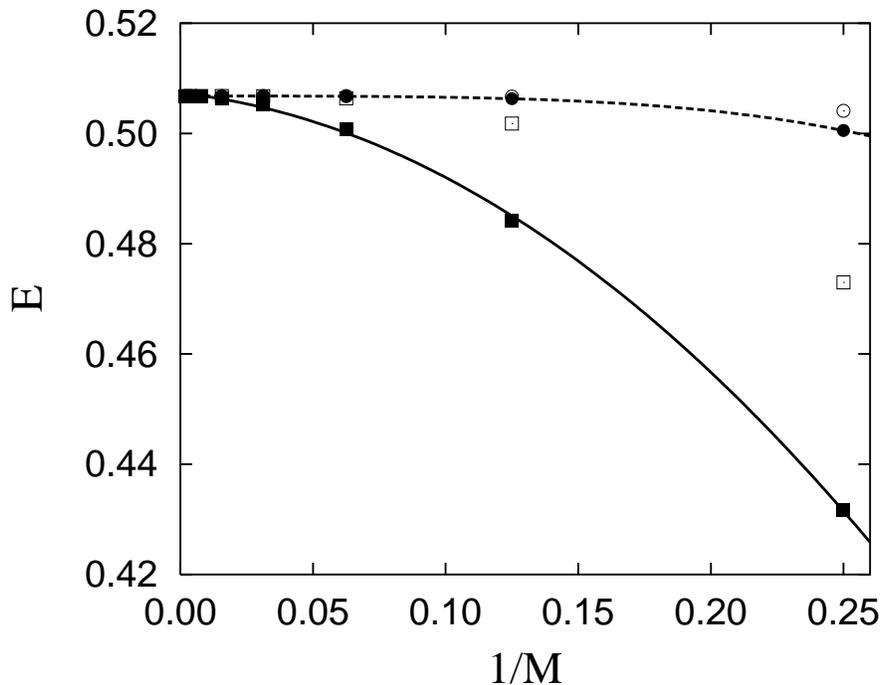}}
\caption{PA (filled squares) and TIA (filled circles) results for the one-dimensional 
harmonic oscillator as a function of $1/M$. Solid and dashed lines are the
fits (\protect\ref{asymtresh}) to the data. Open symbols stand for the
Richardson extrapolations (squares, PA; circles, TIA).} 
\label{harmoniclbpp}
\end{figure}

The dependence of the PIMC results on the \textit{step} $1/M$ is plotted in
Fig. \ref{harmoniclbpp}. As it is expected from the accuracy of the PA and
TIA factorizations, the behavior of the energy when $1/M \rightarrow 0$
follows the power law
\begin{equation}
E= E_0 + A_{\delta} \, (1/M)^\delta \ ,
\label{asymtresh}
\end{equation}
with $E_0$ the exact energy and $\delta=2,4$ for PA and TIA, respectively.
The lines in the Figure correspond to fits  to the PIMC
results when the power law (\ref{asymtresh}) starts to be valid. As
commented before, Richardson extrapolation is a good tool to establish the
smaller value of $M$ from which up PIMC results follow this asymptotic
behavior. On the other hand, one can see in the same Figure the behavior of
the extrapolated energies. The combination of the PA and Richardson
extrapolation reduces dramatically the difference between PA and TIA    
with zero computational cost.

\subsection{\label{sec:neon}Neon at 25.8 K}
The first application to a real system corresponds to a PIMC simulation of
liquid Ne at a experimental point of coordinates: temperature $T=25.8$ K and
density $\rho= 0.0363$ \AA$^{-3}$. We have used a simulation box
containing 108 particles with periodic boundary conditions to represent the
homogeneous liquid. Finite size effects have been studied and proved to be
well corrected by summing up to the energy standard tail corrections to the
potential energy.  The interatomic potential between Ne atoms is the HFD-B
model proposed by Aziz \textit{et al.}\cite{azizne} 

\begin{table}
\centering
\begin{ruledtabular}
\begin{tabular}{ccccccc}
$M$ & $(V/N)_{\textrm{PA}}$ & $ (K/N)_{\textrm{PA}}$ & $(E/N)_{\textrm{PA}}$  &  
 $(V/N)_{\textrm{TIA}}$ & $ (K/N)_{\textrm{TIA}}$ & $(E/N)_{\textrm{TIA}}$ 
\\
\hline 
2 &$-210.25(12)$&$44.42(2)$&$-165.84(14)$&$-207.18(13)$&$48.04(4)$&$-160.14(16)$\\
4 &$-208.30(16)$&$47.80(4)$&$-160.50(19)$&$-207.58(17)$&$49.43(5)$&$-158.15(22)$\\
8 &$-207.80(17)$&$49.05(5)$&$-158.76(22)$&$-207.48(11)$&$49.61(3)$&$-157.87(13)$\\
16&$-207.56(17)$&$49.48(5)$&$-158.08(20)$&$-207.47(20)$&$49.71(6)$&$-157.76(22)$\\
32&$-207.52(22)$&$49.60(6)$&$-157.92(22)$&             &          &             \\
64&$-207.51(44)$&$49.70(9)$&$-157.81(36)$&             &          &             \\
\end{tabular}
\end{ruledtabular}
\caption{Total ($E$), kinetic ($K$), and potential ($V$) energies for a
different number of beads $M$ in liquid Ne at 25.8 K. Figures within brackets
are the statistical errors.}
\label{NeonBisecPP} 
\end{table}

\begin{table}
\centering
\begin{ruledtabular}
\begin{tabular}{ccccccc}
 $M$  & $(E/N)_{\textrm{PA}}$ & $(E/N)_{M}^{(2)}$ & $(E/N)_{\infty}^{(2)}$ &
 $(E/N)_{\textrm{TIA}}$ &  $(E/N)_{M}^{(4)}$ & $(E/N)_{\infty}^{(4)}$ \\   
\hline    
2 &$-165.84(14)$&             &             &$-160.14(16)$&             &             \\
4 &$-160.50(19)$&             &             &$-158.15(22)$&             &  \\
8 &$-158.76(22)$&$-159.16(24)$&$-158.72(25)$&$-157.87(13)$&$-158.03(24)$&$-158.02(24)$\\
16&$-158.08(20)$&$-158.32(28)$&$-158.18(29)$&$-157.76(22)$&$-157.85(14)$&$-157.85(14)$\\
32&$-157.92(22)$&$-157.91(25)$&$-157.85(26)$&		  &$-157.75(24)$&$-157.75(24)$   \\
64&$-157.81(36)$&$-157.88(28)$&$-157.87(29)$&		  &		&	      \\
128&            &$-157.78(45)$&$-157.77(47)$&	       &	     &  	   \\			
\end{tabular}
\end{ruledtabular}
\caption{Richardson extrapolations, to $M$ ($(E/N)_M^{(\delta)}$) and to 
$\infty$ ($(E/N)_{\infty}^{(\delta)}$), 
of the total energy of liquid Ne   with both, PA and TIA.}
\label{Richardsonextrapolationne}
\end{table}

The results obtained using the PA and TIA for the total and partial
energies are reported in Table \ref{NeonBisecPP} for an increasing number
of beads. The value of the kinetic energy, which is compatible with
previous estimations reported in Refs. \onlinecite{peek} and
\onlinecite{glyde}, is a good measure of the relevance of quantum
effects. Subtracting the classical kinetic energy ($3 T/2$) from the PIMC
value, this quantum correction amounts  11 K to be compared with the total
value reported in Table \ref{NeonBisecPP}, 49.7 K . Taking into account the
statistical error bars, the total energy reaches its asymptotic value in terms of $1/M$
for $M=32$ and $M=8$ using the PA and the TIA, respectively. Therefore, 
$M$ is reduced by a factor of four, a figure significantly smaller than the gain
obtained in the harmonic oscillator. In spite of this decrease, it is still
more efficient to use the TIA than the PA since the computation time for
getting the same level of statistical error using TIA is less than the one
required with PA.

\begin{figure}
  \centerline{\includegraphics[width=0.7\linewidth]{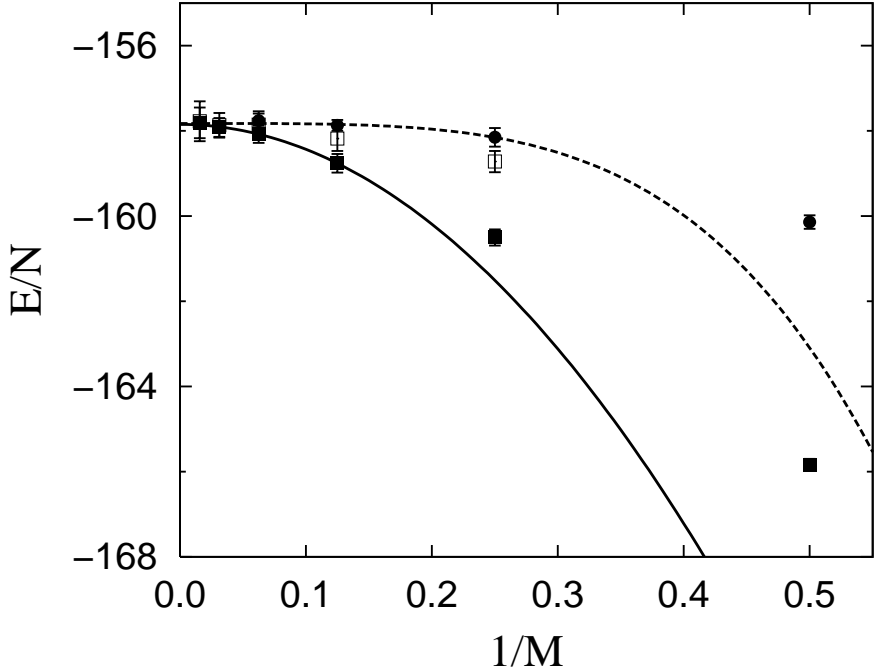}}
\caption{PA (filled squares) and TIA (filled circles) energies for Ne at
25.8 K as a function of $1/M$. Solid and dashed lines are the
fits (\protect\ref{asymtresh}) to the data. Open squares stand for the
Richardson extrapolations using the PA data.}
\label{EnerNeon25}
\end{figure}

The usefulness of the Richardson extrapolation in this system is explored
in Table \ref{Richardsonextrapolationne}. Focusing the analysis on the PA
action, in which the extrapolation is expected to be more useful, one can
see how the extrapolated estimations approach the asymptote faster than 
the direct calculations: starting from  $M=16$ the energy remains constant inside
the confidence interval determined by the statistical error. It is 
obvious that the unavoidable presence of statistical errors in a real
calculation reduces the accuracy of the Richardson extrapolation.
Nevertheless, it always improves the direct calculation and helps in
determining the proximity to the expected power law (\ref{asymtresh}). On
the other hand, in the case of the TIA the extrapolation also improves the
direct estimations but the gain is much smaller than in the PA.

The dependence of the PA and TIA energies on the \textit{step} $1/M$ is shown in 
Fig. \ref{EnerNeon25}. The lines correspond to fits (\ref{asymtresh}) to
the data when $1/M \rightarrow 0$ with $\delta=2,4$ for PA and TIA,
respectively. From the numerical fitting, the final energy per particle is 
 -157.846(20) K for the PA and -157.824(42) K for the TIA. The Figure also
 contains the Richardson extrapolations to $1/M \rightarrow 0$ from  PA,
 which represent a significant improvement with respect the PA data
 and approach the fourth-order behavior of TIA.

\subsection{\label{sec:helium}Liquid $^{\bm 4}$He at 5.1 K}
The third system studied in the present work is the more demanding one:
liquid $^4$He at 5.1 K and density $\rho=0.02186$ \AA$^{-3}$. 
As it is well known, liquid $^4$He is the paradigm of
a quantum Bose liquid and its properties have been well reproduced
theoretically both at zero\cite{boro} and finite
temperature.\cite{ceperley} At 5.1 K, $^4$He is in its
normal phase and the probability of accepting exchanges by introducing the
correct symmetry in the action is very small, so its influence on the total
energy becomes  negligible.\cite{ceperley} Therefore, we consider $^4$He atoms as
Boltzmann-like particles. The interatomic potential is the  HFD-B(HE) model
from Aziz \textit{et al.},\cite{azizhe} used with high accuracy in zero-temperature
calculations, and the homogeneous phase is simulated using a cubic box
containing 64 atoms with periodic boundary conditions. Increasing the
number of atoms, for a fixed number of beads, we have verified that the
tail corrections  added to the energy accounts satisfactorily for
finite-size effects.

\begin{table}
\centering
\begin{ruledtabular}
\begin{tabular}{ccccccccc}
$M$ & $(V/N)_{\textrm{PA}}$ & $ (K/N)_{\textrm{PA}}$ & $(E/N)_{\textrm{PA}}$  &  
 $t_{\textrm{PA}}$ & $(V/N)_{\textrm{TIA}}$ & $ (K/N)_{\textrm{TIA}}$ & 
 $(E/N)_{\textrm{TIA}}$ &   $t_{\textrm{TIA}}$
\\
\hline 
8 &$-23.955(15)$&14.663(15)&$-9.292(20)$& 0.8
&$-22.803(10)$&16.503(25)&$-6.300(26)$ & 2.7\\
16&$-22.737(10)$&16.885(20)&$-5.852(23)$& 1.5
&$-22.124(12)$&18.157(25)&$-3.967(26)$ & 4.5\\
32&$-22.162(13)$&18.180(23)&$-3.982(27)$& 3.3
&$-21.895(12)$&18.852(26)&$-3.043(27)$ & 9.9 \\
64&$-21.940(16)$&18.754(23)&$-3.186(25)$& 9.2
&$-21.846(14)$&18.996(26)&$-2.850(27)$& 26.1\\ 	
128&$-21.870(18)$&18.951(24)&$-2.919(26)$& 17.1
&$-21.850(14)$&19.027(27)&$-2.823(28)$ & 51.9 \\
256&$-21.848(17)$&19.006(23)&$-2.842(25)$& 40.5
&            &          &      &    \\
512&$-21.843(20)$&19.036(30)&$-2.807(32)$& 82.4
&            &          &       &   \\
\end{tabular}
\end{ruledtabular}
\caption{Total ($E$), kinetic ($K$), and potential ($V$) energies for a
different number of beads $M$ in liquid $^4$He at 5.1 K.  $t_{\textrm{PA}}$
($t_{\textrm{TIA}}$) is the CPU time in thousands of seconds of a
PIV-2.4GHz computer using the PA (TIA) approximation.}
\label{HeliumBisecPP} 
\end{table}

\begin{table}[]
\centering
\begin{ruledtabular}
\begin{tabular}{ccccccc}
 $M$  & $(E/N)_{\textrm{PA}}$ & $(E/N)_{M}^{(2)}$ & $(E/N)_{\infty}^{(2)}$ &
 $(E/N)_{\text{TIA}}$ &  $(E/N)_{M}^{(4)}$ & $(E/N)_{\infty}^{(4)}$ \\   
\hline   
8 &$-9.292(20)$&             &            &$-6.300(26)$&                 \\   
16&$-5.852(23)$&             &            &$-3.967(26)$&	    &  \\  
32&$-3.982(27)$&$-4.992(29)$ &$-4.705(31)$&$-3.043(27)$&$-3.821(28)$&$-3.811(28)$\\
64&$-3.186(25)$&$-3.514(34)$ &$-3.359(36)$&$-2.850(27)$&$-2.985(29)$&$-2.981(29)$\\
128&$-2.919(26)$&$-2.987(31)$&$-2.921(33)$&$-2.823(28)$&$-2.838(29)$&$-2.837(29)$\\
256&$-2.842(25)$&$-2.852(33)$&$-2.830(35)$&	       &$-2.821(30)$&$-2.821(30)$   \\
512&$-2.807(32)$&$-2.823(31)$&$-2.816(33)$&	      & 	   &	      \\
1024&           &$-2.798(40)$&$-2.795(43)$	      & 	  &	       &      \\     
\end{tabular}
\end{ruledtabular}
\caption{Richardson extrapolations, to $M$ ($(E/N)_M^{(\delta)}$) and to 
$\infty$ ($(E/N)_{\infty}^{(\delta)}$), 
of the total energy of liquid $^4$He  with both, PA and TIA.}
\label{Richardsonextrapolationhe}
\end{table}

\begin{figure}
   \centerline{\includegraphics[width=0.6\linewidth]{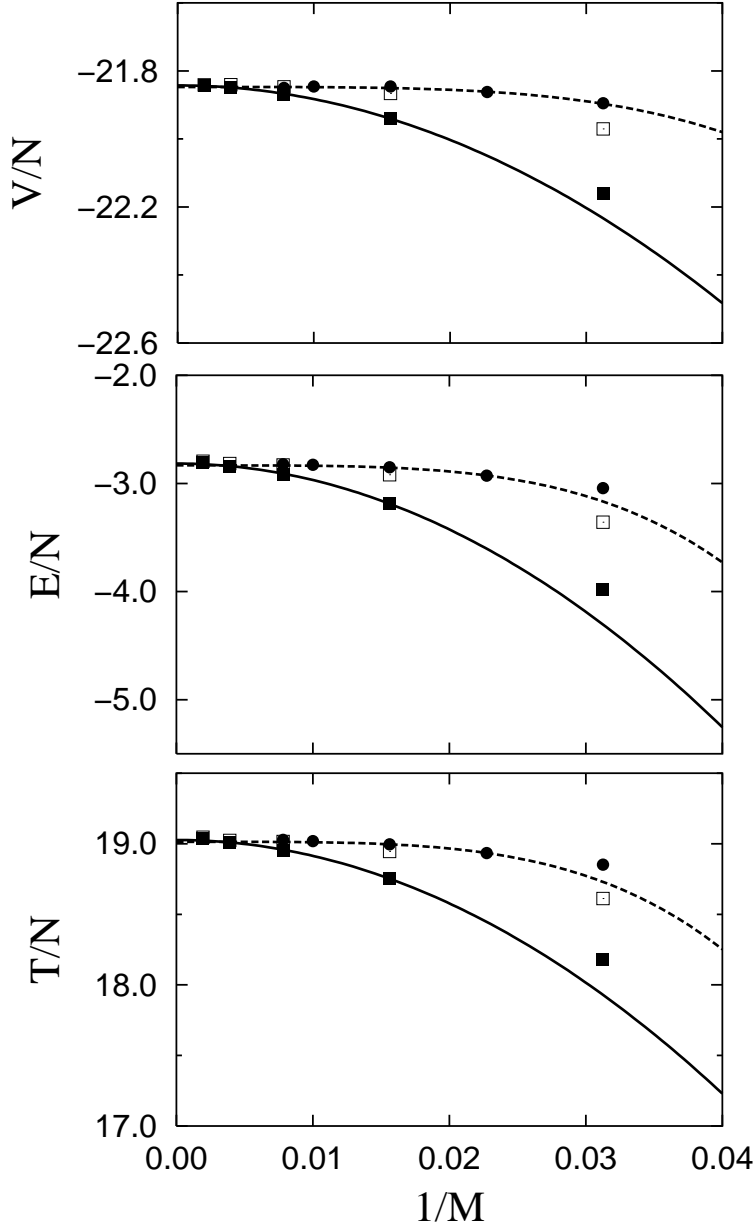}}
\caption{ Potential, total, and kinetic energies of liquid $^4$He at 5.1 K.
PA and TIA data are shown as filled squares and filled circles,
respectively. Solid and dashed lines are the
fits (\protect\ref{asymtresh}) to the data. Open squares stand for the
Richardson extrapolations using the PA data.}
\label{EnerHelium5-2}
\end{figure}

Table \ref{HeliumBisecPP} contains results for the total and partial energies 
for an increasing number of beads $M$. The essentially quantum nature of
$^4$He is reflected in the high value of its kinetic energy, comparable to
the potential energy. The quantum correction on top of the classical value
is $\sim 11.5$ K, to be compared with a total value of 19.0 K. Regarding
the behavior of the PA and TIA energies with $M$, one can observe a similar
behavior to the one obtained for Ne in the previous Subsection. The
asymptotic value using PA is reached for $M=256$ whereas $M=64$ for the
TIA; the accuracy is then improved by  a factor of four. Considering
altogether, the reduction in the number of beads and the additional
computation required for the TIA,  the efficiency of the TIA is still
greater than the one of the PA.

Richardson extrapolations using the PIMC results for the total energies are
reported in Table \ref{Richardsonextrapolationhe}. The main features are
common to the ones observed in Ne. The use of Richardson extrapolation
always improves the direct calculation with zero computational cost and it
arrives faster to the asymptote. From the results contained in the Table,
one can see that the right energy using PA is obtained with data up to
$M=128$, a factor of two smaller than the direct PA calculation. In the case
of the TIA, the extrapolation is also better than the direct output but it
does not significantly improve the efficiency in achieving the plateau.

Figure \ref{EnerHelium5-2} is a plot of the total and partial energies as a
function of $1/M$ near the region where the power laws (\ref{asymtresh})
are satisfied. As it can be seen, the TIA data approaches the limit according
to a fourth-order law, a point that sometimes has been questioned for
hard-core-like interactions at very low temperatures. The Richardson
extrapolations make a good job for both the total and partial energies by
improving significantly the $1/M$ dependence of the PA data. An unbiased 
estimation of the energies is drawn from the 
numerical fits  (\ref{asymtresh}) with $\delta=2$ for PA and 4 for TIA.
Using PA, $K/N=19.026(7)$ K, $V/N=-21.842(2)$ K, and $E/N=-2.816(8)$ K. With
the TIA, $K/N=19.019(6)$ K, $V/N=-21.847(3)$ K, and $E/N=-2.827(5)$ K.     
At the same thermodynamic point, Ceperley and Pollock\cite{pollock2} reported a total
energy of -2.61 K and a kinetic energy of 18.09 K, but using a different
interatomic potential.

\section{\label{sec:conclusions}Conclusions}
In the present work we have explored two possible strategies for improving
the number-of-beads-dependence of the PIMC algorithm, specially for exigent
problems in which the use of PA may require a very large number $M$. First,
the usefulness of the Richardson extrapolation on top of the PA data has
been analyzed for the first time. Richardson extrapolation is a clever way
of practically improving the order of the errors introduced by
discretization and its use in numerical algorithms like integration or
solution of differential equations is widely known. Also in PIMC 
a discretization through an effective step $1/M$ is introduced and one is
interested in going to the limit $1/M \rightarrow 0$. The results obtained
have proved that also in PIMC Richardson extrapolation can be useful since
it approaches faster to the asymptote and helps to know when the discrete
results behave according to the expected power law. It is worth mentioning
that the presence of statistical errors mask somewhat the signal but it is
also true that the computational cost of the extrapolation is certainly
zero.

The second main focus of the work has been to study the behavior of the TIA
in systems with hard-core-like interactions and well into the quantum
regime. Previous tests of the TIA\cite{muser,lacks,weht,singer}
were performed in less exigent conditions
and some doubts concerning the fail of the
predicted fourth-order accuracy were formulated. The present results, mainly the ones
achieved in the calculation of $^4$He, have shown that the fourth order law
is maintained. On the other hand, the TIA introduces in the
calculation additional computation due to the need of the second derivative of the
potential (using the virial estimator for the kinetic energy, absolutely
necessary when $M$ increases). This is not a serious problem, at least
when empirical potentials are used, and the computer time to achieve the
same statistical error in the asymptotic limit in $M$ is ever reduced. Our
present experience shows that the CPU time is at least a 30\% smaller, and that this
percentage increases with the number of beads required to reach the
asymptote.

As a final point, the fact that the double commutator $[[V,K],V]$  in the TIA has
allowed for a substantial decrease of the systematic errors for very
different potentials is an encouraging result. At present,
Chin\cite{chin2} is developing various families of symplectic
algorithms  in which that double commutator
originates a fourth-order correction which can be fine-tuned
both in sign and magnitude. This bears the promise of achieving an
effective cancellation of the leading fourth and sixth order corrections,
feature which could result in a dramatic reduction of the computational
cost of PIMC calculations. Further work in this direction would be most
interesting.

\acknowledgments
This work was supported, in part, by the DGI (Spain) Grant No.
BFM2002-00466, Generalitat de Catalunya Grant No. 2001SGR-00222, 
and project MIUR-2001/025/498 (Italy).


\end{document}